# Creation of Ghost Illusions Using Metamaterials in Wave Dynamics

By *Wei Xiang Jiang*, *Cheng-Wei Qiu\**, *Tiancheng Han*, *Shuang Zhang, and Tie Jun Cui\**
 (W.X.J. and C.W.Q. contributed equally)

Prof. Tie Jun Cui, Dr. Wei Xiang Jiang

State Key Laboratory of Millimeter Waves, Department of Radio Engineering

Southeast University, Nanjing 210096 (China)

E-mail: tjcui@seu.edu.cn

Prof. Cheng-Wei Qiu, Dr. Tiancheng Han

Department of Electrical and Computer Engineering, National University of Singapore

119620 (Singapore)

E-mail: chengwei.qiu@nus.edu.sg

Prof. Shuang Zhang

School of Physics and Astronomy, University of Birmingham, Edgbaston

Birmingham, B15 2TT (UK)

**Abstract:** The creation of wave-dynamic illusion functionality is of great interests to various scientific communities, which can potentially transform an actual perception into the pre-controlled perception, thus empowering unprecedented applications in the advanced-material science, camouflage, cloaking, optical and/or microwave cognition, and defense security, etc. By using the space transformation theory and engineering capability of metamaterials, we propose and realize a functional "ghost" illusion device, which is capable of creating wave-dynamic virtual ghost images off the original object's position under the illumination of electromagnetic waves. The scattering signature of the object is thus ghosted and perceived as multiple ghost targets with different geometries and compositions. The ghost-illusion material, being inhomogeneous and anisotropic, was realized by thousands of varying unit cells working at non-resonance. The experimental demonstration of the ghost illusion validates our theory of scattering metamorphosis and opens a novel avenue to the wave-dynamic illusion, cognitive deception, manipulate strange light or matter behaviors, and design novel optical and microwave devices.

## 1. Introduction

Transformation optics, a new designing tool based on space transformation, has enabled a number of unconventional optical devices, including perfect invisibility cloaks [1-4], carpet cloaks [5-10], illusion-optics devices [11-15], Eaton lens [16], and flattened Luneburg lenses [17,18], etc. In particular, the creation of optical illusion has been of great interest since it may lead to many unprecedented applications [11-15]. Illusion optics was proposed based on a two-fold operation [11]. The first is to eliminate the scattering of the original object by using double-negative metamterial, and the second is to creat the desired scattering signature by employing additional spatial transformations. However, a number of implications are entailed by involving lossless double-negative metamaterial in the transformation. For instance, from theoretical point of view, not all objects have their complementary counterparts (e.g., perfect electric and magnetic conductors, gyrotropic objects or chiral objects, etc). In the mean time, it is too stringent in practice, if not impossible, to achieve lossless double-negative materials in electromagnetic and optical spectra. Hence, the loss will dramatically deteriorate the illusional performance. In this connection, an improved illusion device consisting of only positive materials has been studied [12]. However, extreme values of optical parameter are required, which would necessarily involve resonance effect that leads to narrow band operation, not to mention the complexity of metamaterial designs.

Double-negative metamaterials may be easily obtained in composite transmission-line RLC circuits. Though the planar RLC circuits can exhibit the superscattering effect via the manipulations of voltage and impedance, it is inherently restricted to lumped circuitry on a circuit board, and thus cannot be considered as illusion per se, based on light-matter interaction in free space governed by Maxwell's equations. Beyond the circuit representation, two spatial wave-dynamic experiments have been demonstrated to manifest the real radar illusions [14,15]. However, such experiments have critical restrictions: they can only simply reduce the size or change the material property of the original object, but cannot alter and shift the scattering center of the original object. That is to say, there is only one single illusion object at the original position. Therefore, the observer can still perceive that there is one scattering center at the position of the object, though with slightly different scattering signatures.

In this paper, we report a distinguished anisotropic-graded-material (AGM) based functional illusion device in the wave dynamics, which virtually transforms the scattering signature of an arbitrary object to that of two or more (multiple) isolated ghost objects (i.e., not existing physically). In other words, the ghost illusion device can make the scattering signature of one object equivalent to that of two or more other objects, which are arbitrarily pre-designed, not only in the near-field scattering but also in the far-field scattering cross section.

As we have stressed, the original illusion optics needs negative-index complementary media in a two-fold process (to cancel the old scattering and then create a new one), but we only used anisotropic graded materials in one-fold process, i.e., directly metamorphosing the scattering to the pre-controlled pattern. Moreover, the theory of optical illusion only presented to change the scattering metric of one to another, but the scattering center is still at the original center. On the contrary, our theory and experiment unanimously demonstrate multiple ghost scattering centers off the original onject. Hence the proposed ghost illusion device behaves in a distinguished manner from the functionality of optical illusions.

## 2. Functional Ghost Illusion Device

### 2.1. Concept and Analytical Design

The functions of the ghost illusion device are schematically illustrated in **Figure 1**, which transforms a perfectly conducting (or metallic) object in Figure 1(a) into three objects, including a shrunk metallic object in the original position and two "virtual" dielectric objects (which can also be metallic) on each side, as in Figure 1(c). In comparison, the radar signature of the original metallic object is given in Figure 1(d), showing the typical feature of a single target. When the metallic object is covered by the designed ghost illusion device (see Figure 1(b)), the corresponding radar signature is demonstrated in Figure 1(e), which is completely different from Figure 1(d), but exactly the same as that of three virtual objects (Figure 1(c)) shown in Figure 1(f). Hence the ghost device shrinks the original metallic object and produces two additional ghost images in the radar signature. In other words, radar only perceives that there are three standalone objects, though only a single object (the real object with the ghost coating) is physically present at the center.

The above functionality can be directly derived using the transformation optics theory [2]. In the actual space, the radius of the object (or the inner radius of the ghost device) is *a*, and the outer radius of the ghost device is *c*. In the virtual space, there are three distinct objects: a shrunk object and two wing-ghosts, as shown in Figure 1(c). Here, we just design two wing-ghosts as an example. Actually, we can design arbitrary number of virtual objects, so that it is more rotationally symmetric and is less independent of the incident wave. The radius of shrunk object is *a'* and the inner and outer radii of wing-ghosts are *b'* and *c*. The general three-dimensional (3D) optical transformation for the above functionality is written in the spherical coordinates as

$$(r,\theta,\varphi) = (k(r'-a')+a, \theta', \varphi') \qquad (1)$$

in which $k = \dfrac{c-a}{c-a'}$. Then the constitutive parameters in the region from $r = a$ to $r = c$ except the two wing-ghost areas (see region I in **Figure 2**(b)) and the two wing-ghost areas (region II in Figure 2(b)) in the spherical coordinates can be obtained as

$$(\varepsilon_r, \varepsilon_\theta, \varepsilon_\varphi) = (\mu_r, \mu_\theta, \mu_\varphi) = \left(\dfrac{1}{k}\left(\dfrac{r-a+ka'}{r}\right)^2, \dfrac{1}{k}, \dfrac{1}{k}\right) \quad \text{(for region I)} \qquad (2a)$$

$$(\varepsilon_r, \varepsilon_\theta, \varepsilon_\varphi) = (\mu_r, \mu_\theta, \mu_\varphi) = \left(\dfrac{1}{k}\left(\dfrac{r-a+ka'}{r}\right)^2, \dfrac{1}{k}, \dfrac{\varepsilon_{virtual}}{k}\right) \quad \text{(for region II)} \qquad (2b)$$

where, $(\varepsilon_r, \varepsilon_\theta, \varepsilon_\varphi)$ and $(\mu_r, \mu_\theta, \mu_\varphi)$ are the relative permittivity and permeability tensors, and $\varepsilon_{virtual}$ is the relative permittivity of the ghost-wing targets, which is a pre-designed and known parameter. Equation (2) describes the electromagnetic properties of a 3D ghost-illusion device. For two-dimensional (2D) case, the cross section of the cylindrical ghost device is the same as in Fig. 1. The cylindrical device is characterized by its diagonal-parameter tensors in the cylindrical coordinates $(r, \varphi, z)$, and the same equations shown above hold for the diagonal-components ($\mu_r, \mu_\varphi, \mu_z, \varepsilon_r, \varepsilon_\varphi, \varepsilon_z$) of full parameters. In order to be realized by artificial materials, the full parameter can be simplified under the transverse-electric (TE) polarization. The simplified parameters of the 2D ghost-illusion device for $z$-polarized electric fields are written as

$$(\mu_r, \mu_\varphi, \varepsilon_z) = \left(\left(\dfrac{r-a+ka'}{r}\right)^2, 1, \dfrac{1}{k^2}\right) \quad \text{(for region I)} \qquad (3a)$$

$$(\mu_r, \mu_\varphi, \varepsilon_z) = \left(\left(\dfrac{r-a+ka'}{r}\right)^2, 1, \dfrac{\varepsilon_{virtual}}{k^2}\right) \quad \text{(for region II)} \qquad (3b)$$

We illustrate the above medium parameters of the ghost device in regions I and II in Figure 2(a) as a function of the radian variable $r$, in which the geometrical parameters are set by $a'=0.41\lambda_0$, $a=0.95\lambda_0$, $b'=1.08\lambda_0$, $c=1.75\lambda_0$, $\varepsilon_{virtual}=2.23$, and the free-space wavelength $\lambda_0=30$mm at the frequency of 10 GHz. From Figure 2(a), we notice that the relative permittivity in two regions are 2.82 and 6.29, respectively, and the relative permeability ranges from 0.06 to 0.35, which can be realized by using artificial materials.

**2.2. Material Synthesis and Device Fabrication**

In region I, the permittivity maintains 2.82 and the permeability ranges from 0.065 to 0.226, which are realized by artificially structured materials – the conventional split-ring resonators (SRRs) etched on a dielectric substrate; while in region II, the permittivity maintains 6.29 and the permeability ranges from 0.226 to 0.355, which are realized by a class of modified SRRs shown in Figure 2(b). Such modified SRR structures can raise the permittivity remarkably. In our design, there are two methods to generate the required high electric permittivity (6.29) in Region II without using any auxiliary high-permittivity materials. One is to add some curved structures to SRRs, and the other is to reduce the wire width of SRRs.

The designed ghost device illustrated in Figure 2(b) is synthesized with eight concentric layers of low-loss printed circuit boards (PCBs). On each layer, SRRs of 35μm-thick copper was coated on one side of the 0.25mm-thick substrate (F4B) with the relative permittivity $\varepsilon=2.65+i0.003$. Although we apply different structures in Regions I and II, the height of each layer ($0.36\lambda_0$, 10.8 mm) and the distance between adjacent layers ($\lambda_0/10$, 3 mm) remain the same. The inner and outer radii of the ghost device are $0.95\lambda_0$ and $1.75\lambda_0$, respectively. Each layer carries three rows of SRRs, whose length and height are 3.14 mm and 3.6 mm. The designed SRRs create the required radial permeability $\mu_r$ and $z$-direction permittivity $\varepsilon_z$ at 10 GHz. The PCB rings are adhered to a 2-mm-thick hard-foam board that was cut with concentric circular rabbets to fit the concentric layers exactly.

**2.3. Full-Wave Simulation Results**

The performance of the ghost device is numerically verified at the operating frequency of 10 GHz ($\lambda_0$=30mm), in which the geometrical parameters are chosen to be the same as those in later experiments: $a'=0.41\lambda_0$, $a=0.95\lambda_0$, $b'=1.08\lambda_0$, $c=1.75\lambda_0$, and $\varepsilon_{virtual}=2.23$. The distance between the source and the center of ghost device is $2.43\lambda_0$. The numerical simulation results are presented in **Figure 3**, in which (a) and (c) illustrate the electric-field distributions of a metallic cylinder without and with the full-parameter ghost device, respectively, under the excitation of a line source.

Figure 3(b) shows the electric fields of a small metallic cylinder with two standalone dielectric wings, which are characterized by the relative permittivity 2.23. The effectiveness of the ghost device in creating radar ghost images is confirmed via the equivalence between the identical patterns in Figures 3(b) and 3(c) outside the outermost boundary. In other words, the proposed ghost device virtually camouflages the initial radar image of a metallic cylinder by illusion radar images of three objects. More interestingly, two ghost images appear separately apart from the original position, which totally change the signature of the given target. In order to facilitate the fabrication in experiments, it is necessary to consider and check the performance of the simplified ghost device, whose electric-field patterns are illustrated in Figure 3(d). It can be seen that the radar signature of the metallic cylinder coated with the simplified ghost device is nearly the same as those in Figures 3(b) and 3(c). Hence it is justified to use the simplified parameters as the experimental verification of the ghost effects.

The ghost device is also effective for dielectric objects. For example, we can simply choose the object as free space (or air) with the same geometrical parameters. The numerical simulation results are demonstrated in Figures 3(e) and 3(f), from which we observe that such a ghost device can virtually transform the void (i.e., air) into the electromagnetic images of three distinct objects. It is noted that the ghost device not only shrinks the central air area, but also changes the material properties virtually.

## 2.4. Experimental Results

**Figure 4**(a) illustrates the experimental setup together with the resultant 2D fields in the near zone. The electric field depends on *x* and *y* coordinates, but does not vary in the *z* direction, which is the direction of polarization parallel to the antenna's orientation. In the experiment, the distance between the source and the center of ghost device is set to be $3.93\lambda_0$. Figures 4(b)-(d) show the simulation and measurement results of the ghost device. As illustrated in Figure 4(b), the measured electric fields inside the ghost device are perturbed by AGM, but the scattered waves outside are in very good agreement to the numerical results for a layered ghost-device profile, as shown in Figure 4(d). Figure 4(c) compares the simulated and measured electric field intensity along a pre-selected line, located at $3.8\lambda_0$ (114 mm) away from the ghost device. In this experiment, the ghost device has just 8 layers including two regions as Figure 2(b) shows, and we are already able to observe good illusion performance. The simplified prototype preserves the functionality of the ideal ghost device, indicating the robustness of the design for practical applications.

## 3. Extending Functionalities

In the above simulation and experiment results, we have discussed and validated one situation of the ghost device, which transforms a metallic object in the real space into a smaller metallic object with two standalone dielectric wings in the virtual space. It is noted more types of ghost scattering can be produced under other design configurations. For example, when we choose *a'*=0 mm (i.e., the radius of metallic cylinder is zero in the virtual space) with other parameters unchanged, then the device will produce only two ghost-wing objects. The full-wave simulation results of electric fields are presented in **Figures 5**(a) and (b), in which (a) indicates the scattering pattern of two standalone dielectric wings with relative permittivity 2.23, and (b) illustrates the electric-field distribution of a copper cylinder covered by the ghost device. The effectiveness of the ghost device in creating the illusion is obviously pronounced via the identical patterns in these two subfigures outside the outer boundary. In other words, this ghost device virtually camouflages the initial electromagnetic scattering of a metallic cylinder, as if there were only those of two dielectric objects. It is very important that there will be no scattering center in the original position of the actual target in this case.

When the two wings of ghost device are metallic, they are also effective for the dielectric and metallic objects. Figures 5(c)-(f) illustrate two cases of ghost devices with metallic wings. In Figures 5(c) and (d), the ghost device shrinks a copper cylinder to nothing but virtually produces two wing metallic objects; in Figures 5(e) and (f), the shrunk object is dielectric while the two wings are metallic. In both cases, the excellent ghost-illusion performance is clearly observed.

Furthermore, if we divide Region II into several smaller areas or consider more blocks of Region II, the ghost device will transform the signature of one object to an equivalence of a cluster of multiple different objects. Hence, many interesting illusion performance can be manipulated via the proposed ghost devices. Introducing more blocks of Region II can also make the device rotationally symmetric, thus becoming more independent of the incident direction.

## 4. Conclusions

We have conjectured, fabricated, and demonstrated a general ghost-illusion device enabling an object to possess arbitrary wing-ghosts off its original position and produce vitual images. The principle can be precisely controlled for producing the desired scattering metrics of the ghost device. Metamaterial structures using thousands of SRRs are employed to realize the 2D prototype, so that the ghost signatures are produced at non-resonant frequency with low-loss feature thereby. The metamorphosis of electromagnetic signature has been verified by numerical investigation and experiments, which demonstrate good ghost-illusion performance in wave dynamics.

In contrast to cloaking, the ghost devices empowers a novel way to alter the scattering signatures of targets to be "ghosted". The observer cannot distinguish, and the readout can thus be misled. The object is there, but you don't see it while see "them" elsewhere instead. More importantly, the proposed approach can deal with arbitrary target (e.g., perfectly electrical conductor) in wave dynamics without negative-index complementary media, and the ghost is pre-controllable. The current work may provide unprecedented avenue to the light-matter behavior control and security enhancement, which advances beyond the earlier illusion devices [11-15].

## 5. Experimental Section

*Experimental Setup and Field Mapping*: We use a vector network analyzer (Agilent PNA-LN5230C) to excite and receive the microwave signals, and apply a parallel-plate mapping system to scan the electric-field distributions (see **Figure 6**(a)). The electric fields are polarized in the vertical direction confined by two large aluminum plates, whose distance is set as 12 mm. The electric field is detected and scanned by a monopole probe embedded in the top aluminum plate. The feeding and detection probes are connected to two ports of network analyzer via thin coaxial cables. The ghost-device sample is fixed on the center of the bottom plate, which is mounted on a computer-controlled stage that can move in two dimensions. The scanning region covers $4.27\lambda_0 \times 4.67\lambda_0$ ($128 \times 140$ mm$^2$) with a step resolution of 1 mm.

*Fabricated Ghost Device and Experiment Platform*: The photo of fabricated ghost device is shown in Figure 6(b). The detailed values of geometry and material parameters are illustrated in **Table 1**. Throughout the paper, we set $\varphi_0 = 60°$.

**Acknowledgements**

This work was supported in part from the National Science Foundation of China under Grant Nos. 60990320, 60990321, 60990324, 61171024, 61171026, 60901011, and 60921063, in part from the National High Tech (863) Projects under Grant Nos. 2011AA010202 and 2012AA030702, in part from the 111 Project under Grant No. 111-2-05, in part from the Talent Project of Southeast University, and in part by the Joint Research Center on Terahertz Science.

[19] T. C. Han, C.-W. Qiu, and X. H. Tang, *Opt. Lett.* **2010**, 35, 2642.

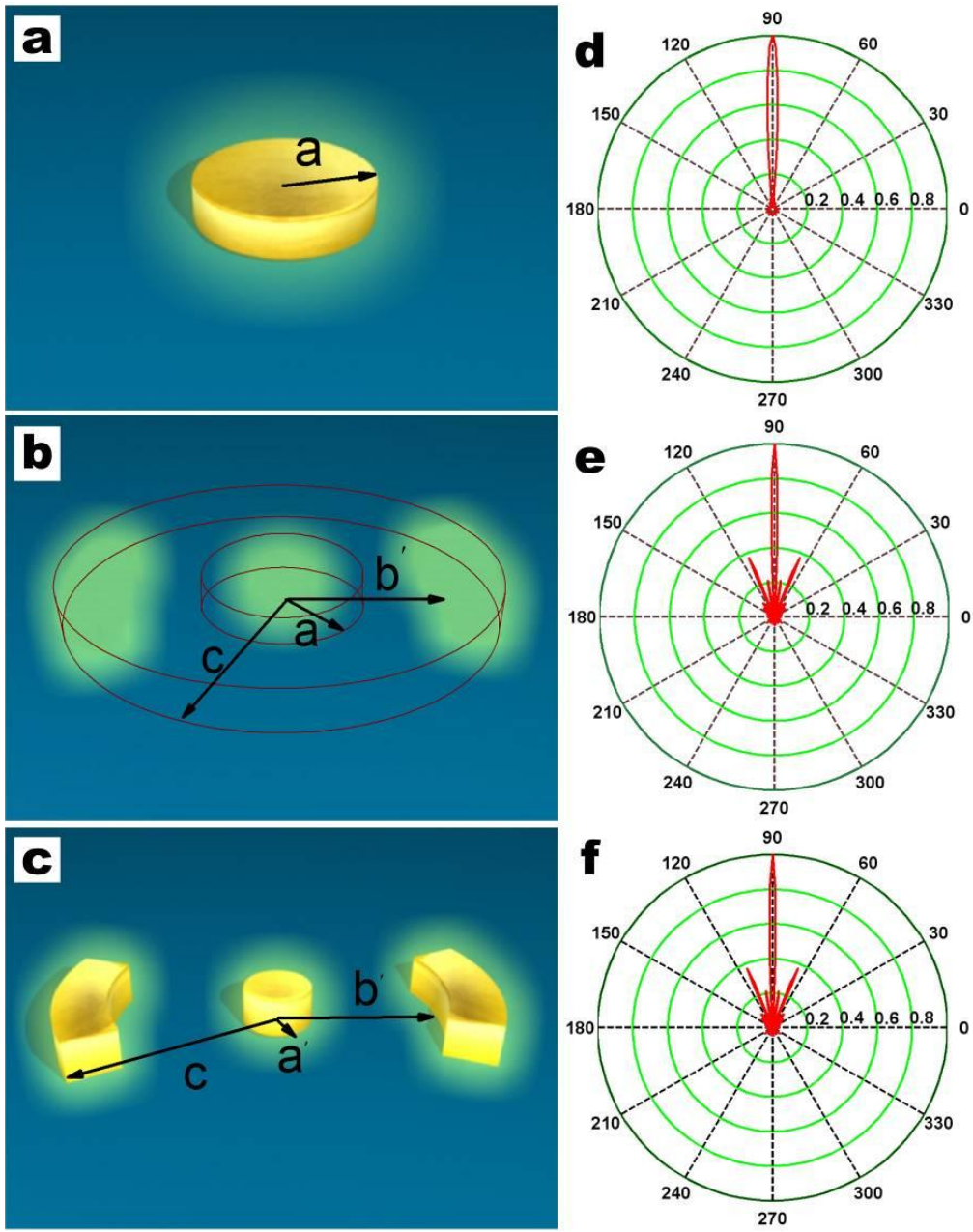

**Figure 1.** Schematic, composition, and equivalence of the ghost-illusion device. (a) The original metallic object with its scattering signature given in panel (d) showing the typical feature of a single target. (b) The metallic object covered by the designed ghost device whose physical boundaries are denoted with red solid lines and scattering signature is given in panel (e), metamorphosing scattering feature of the original object. (c) A shrunk metallic object at the original center with two wing dielectric objects, whose signature is given in panel (f), showing the typical scattering feature of three objects. The scattering signatures of (b) and (c) are completely the same. In the far-field calculation, the geometrical parameters are chosen as $a=3.8\lambda_0$, $a'=1.63\lambda_0$, $b'=5.4\lambda_0$ and $c=7\lambda_0$.

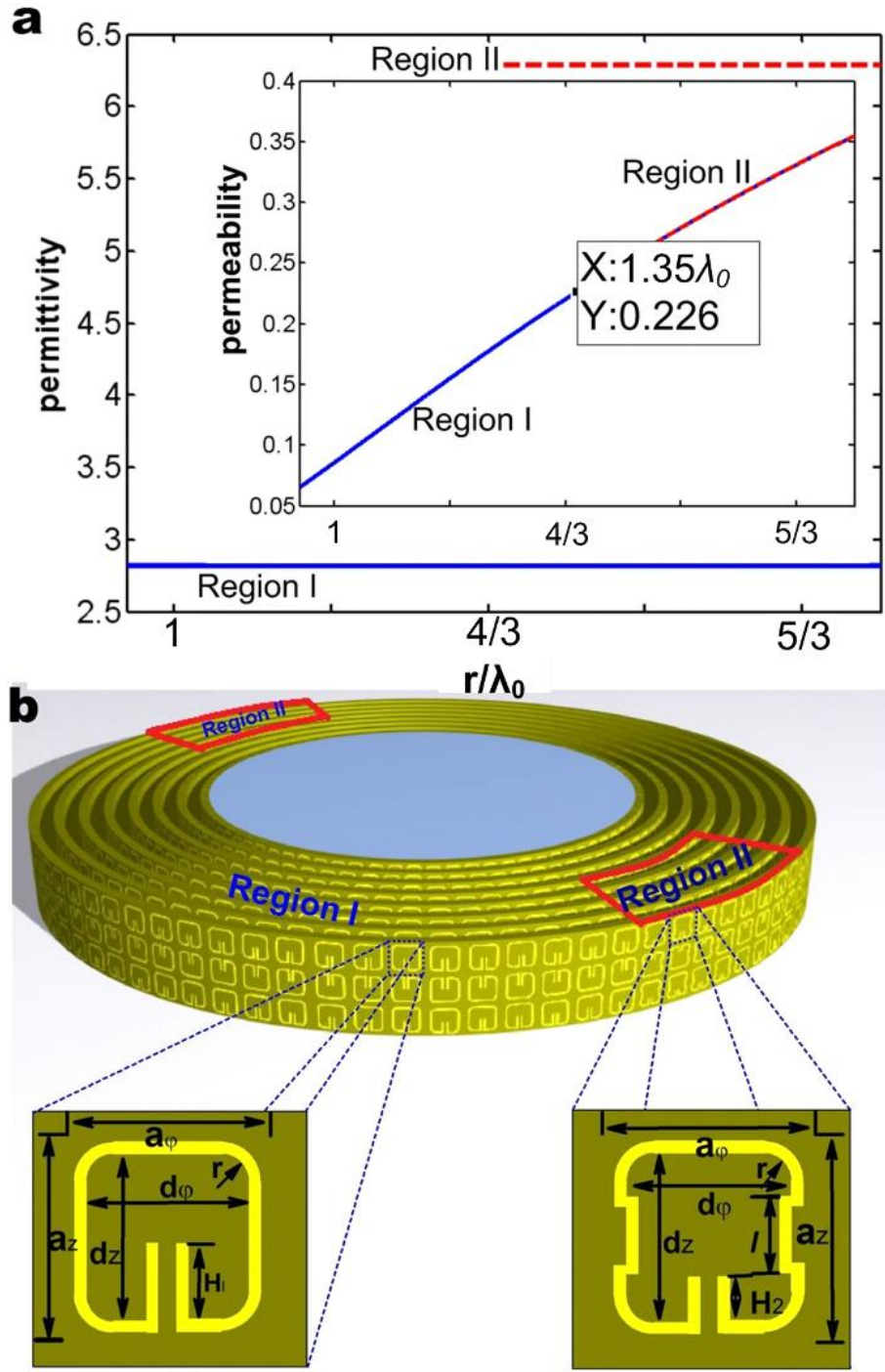

**Figure 2.** The design and fabrication of the simplified ghost device. (a) The radial permeability and *z*-direction permittivity in Regions I and II as a function of radial variable *r*. (b) The conventional and modified SRRs constitute the ghost device following the parameters described in Eq. (2).

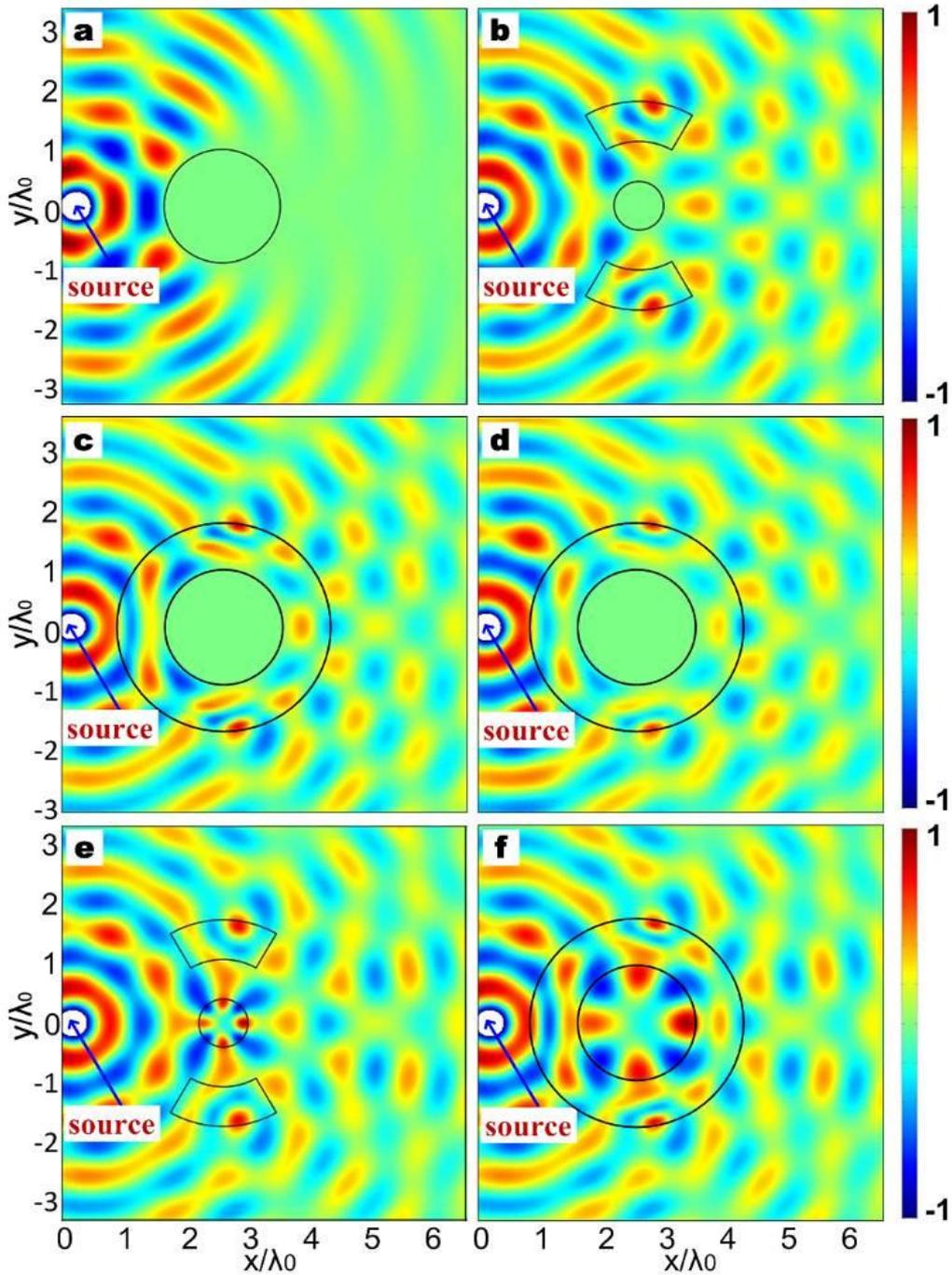

**Figure 3.** Full-wave simulation results of electric-field distributions for different cases. (a) A bare metallic cylinder, whose scattering signature has a big shadow in the forward direction. (b) The virtual objects: a smaller metallic cylinder in the center with two dielectric wing objects, whose scattering signature is completely metamorphosed. (c) The metallic cylinder wrapped by full-parameter ghost device, whose signature is identical to that of three virtual objects. (d) The metallic cylinder wrapped by the reduced-parameter ghost device, whose scattering signature is nearly the same as that of three virtual objects. (e) The virtual objects: a smaller dielectric cylinder in the center with two dielectric wing objects. (f) The dielectric cylinder wrapped by the full-parameter ghost device, whose signature is identical to that of three virtual objects.

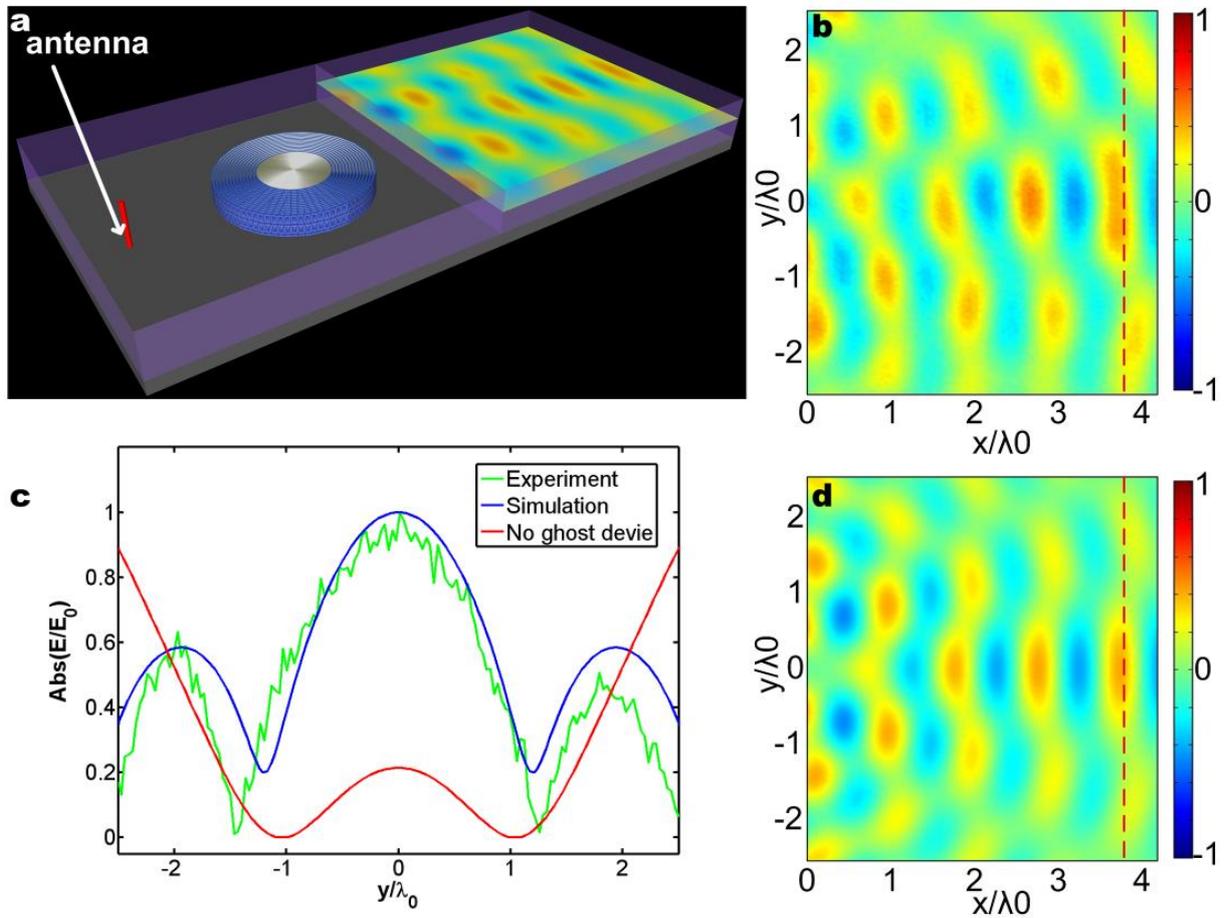

**Figure 4.** The experimental setup, simulation and experimental results of the ghost-illusion device. (a) Illustration of the experimental setup. (b) The measurement result. (c) The normalized intensity along the red dashed line $x=3.8\lambda_0$. (d) The simulation result. The simulation and experiment results have good agreements, demonstrating the ghosting phenomena in scattering signatures.

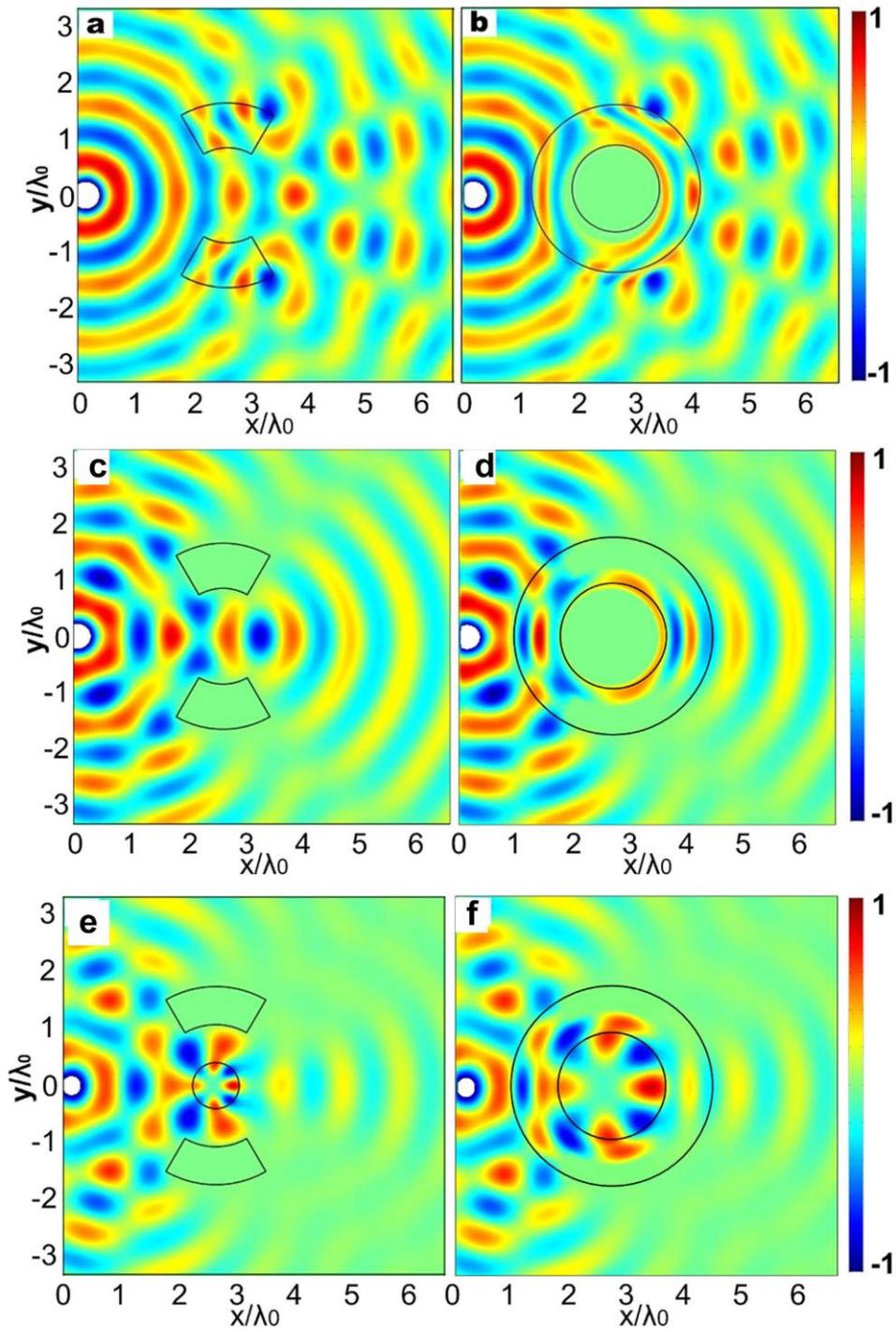

**Figure 5.** Simulation results of other functional ghost-illusion devices. (a) and (b): A ghost device shrinks a metallic cylinder to nothing and virtually produces two wing dielectric objects. (c) and (d): A ghost device shrinks a metallic cylinder to nothing and virtually produces two wing metallic objects. (e) and (f): A ghost device shrinks a dielectric object and virtually produces two wing metallic objects.

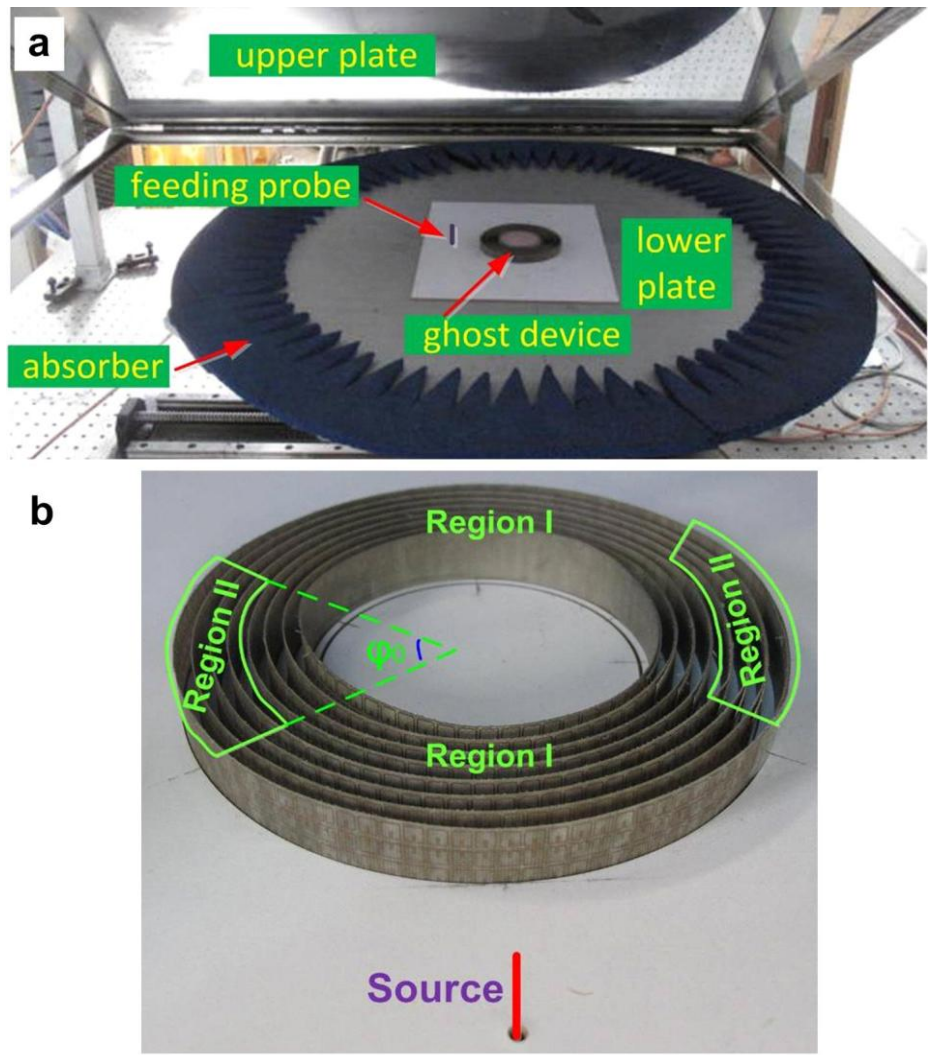

**Figure 6.** (a) The photograph of the parallel-plate waveguide mapping system. (b) The photograph of the fabricated ghost-illusion device.

**Table 1.** The geometrical parameters $h_1$ and $h_2$ of the ghost device, and the associated values of $\varepsilon_z$ and $\mu_r$. The imaginary parts of the electromagnetic parameters are small enough to be neglected.

| No. of Layers | Region I | | | Region II | | |
|---|---|---|---|---|---|---|
| | $h_1$ (mm) | $\varepsilon_z$ | $\mu_r$ | $h_2$ (mm) | $\varepsilon_z$ | $\mu_r$ |
| 1 | 0.894 | 2.847 | 0.086 | | | |
| 2 | 0.939 | 2.835 | 0.127 | | | |
| 3 | 0.986 | 2.824 | 0.168 | | | |
| 4 | 1.057 | 2.810 | 0.218 | | | |
| 5 | 1.097 | 2.803 | 0.244 | 0.618 | 6.253 | 0.244 |
| 6 | 1.162 | 2.793 | 0.279 | 0.776 | 6.229 | 0.279 |
| 7 | 1.228 | 2.785 | 0.311 | 0.957 | 6.207 | 0.311 |
| 8 | 1.298 | 2.776 | 0.341 | 1.169 | 6.185 | 0.341 |

Keyword: Illusion, ghost device, transformation optics, metamaterials, wave dynamics

Wei Xiang Jiang, Cheng-Wei Qiu*, Tiancheng Han, Shuang Zhang and Tie Jun Cui*

Title: Creation of Ghost Illusions Using Metamaterials in Wave Dynamics

ToC figure

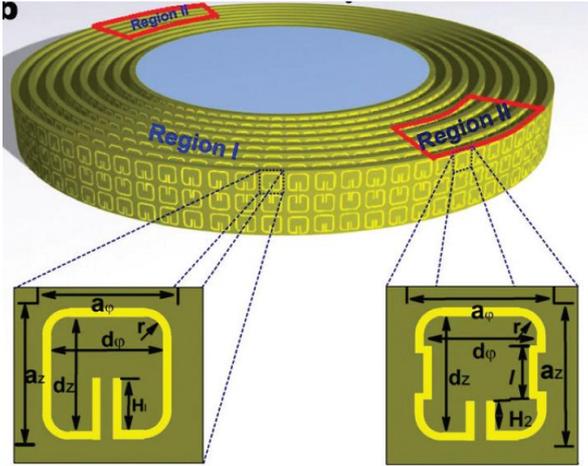